\documentclass{article}
\usepackage{amsmath,amssymb,mathrsfs,epsfig,wick,feyn,hyperref}
\newcommand{\uy}{\underline{y}}

\newcommand{\bx}{\boldsymbol{x}}
\newcommand{\by}{\boldsymbol{y}}

\newcommand{\bgamma}{\vec{\gamma}}
\newcommand{\wickdots}[1]{\protect{\protect{\boldsymbol{:}}#1%
    \protect{\boldsymbol{:}}}}
\newcommand{\Langle}{\big\langle}
\newcommand{\Rangle}{\big\rangle}

\newcommand{\dotsw}{{\protect{\cdot\cdot\cdot\;}}}

\newcommand{\Pf}{\scriptscriptstyle +} 
\newcommand{\Nf}{\scriptscriptstyle -} 
\newcommand{\PNf}{\scriptscriptstyle \pm} 

\numberwithin{equation}{section}

\title{Non Local Theories: New Rules for Old Diagrams}

\author{Gherardo Piacitelli\thanks{Research partially supported by 
MIUR and GNAMPA-INdAM.}\\ \small Dipartimento 
di Matematica, Universit\`a di Roma ``La Sapienza'',\\
 \small P.le Aldo Moro~5, 00185, Roma, Italy.\\
{\small \tt piacitel@mat.uniroma1.it}}

\begin{document}

\maketitle

\begin{abstract}
We show that 
a general variant of the Wick theorems can be used to reduce
the time ordered products in the Gell-Mann \& Low formula
for a certain class on non local quantum field theories, including the
case where the interaction Lagrangian is defined in terms of twisted products.

The only
necessary modification  is the replacement of the Stueckelberg-Feynman
propagator by the general propagator (the ``contractor'' of Denk and Schweda)
\[
\mathscr D(y-y';\tau-\tau')=\frac{1}{i}\bigg(
\Delta_{\Pf}(y-y')\theta(\tau-\tau')+
\Delta_{\Pf}(y'-y)\theta(\tau'-\tau)\bigg),
\]
where the violations of locality and causality are represented by the
dependence of $\tau,\tau'$ on other points, besides those involved in the 
contraction.

This leads naturally to a diagrammatic expansion of the Gell-Mann \& Low
formula, in terms of the same diagrams as in the local case, the only necessary
modification concerning the Feynman rules. The ordinary local theory is easily
recovered as a special case, and there is a one-to-one correspondence 
between the local and non local contributions corresponding to the same 
diagrams, which is preserved while
performing the large scale limit of the theory.
\end{abstract}

\section{Introduction.}

We consider the Hamiltonian setup for perturbations of a
single quantum neutral scalar field~$\phi$ on the Minkowski spacetime, 
with a non local self interaction Lagrangian of the form
\begin{equation}
\label{L}
\mathscr L_I(x)=\int dy_1\dotsm dy_n\;W_x(y_1,\dotsc,y_n)
\wickdots{\phi(y_1)\dotsm\phi(y_n)}.
\end{equation}
If the kernel is chosen to fulfill
\[
W_{x+a}(y_1,\dotsc,y_n)=W_x(y_1-a,\dotsc,y_n-a),
\]
the interaction Hamiltonian 
$
H_I(t)=-g\int d\bx\;\mathscr L_I(t,\bx)
$
is invariant under translations.

Motivations come from the quest for effective theories on classical
spacetime, describing 
interactions on the quantum spacetime, as it was proposed by 
Doplicher, Fredenhagen 
and Roberts (DFR) in their seminal paper~\cite{dfr}; see 
also~\cite{dfrshort}. 
There, localization is described in terms of 
the covariant DFR quantum coordinates~$q^\mu$, 
whose commutators 
$i\lambda_P^2Q^{\mu\nu}=[q^\mu,q^\nu]$ are constrained so to ensure
stability of spacetime under localization ($\lambda_P$ is the Planck length):
no black holes
should arise as an effect of localization 
{\em alone}. Note that, for this class of models, 
sharp localization still can be obtained in one coordinate, but not in all
the coordinates simultaneously; pointwise localization is prevented 
by the uncertainty relations among the coordinates. 

In particular, we consider the interaction
\begin{equation}
\label{LQ}
\mathscr L_I^Q(x)=\int_{{\Sigma_1}} 
d\sigma\;\mathscr L_I^\sigma(x),
\end{equation}
where
\[
\tag{1.2'}\label{Lsigma}
\mathscr L_I^\sigma(x)=\wickdots{(\phi\star_\sigma\dotsm\star_\sigma\phi)(x)},
\]
$\star_\sigma$ is the product twisted by the 
real antisymmetric matrix~$\sigma$, and~$d\sigma$ is the rotation
invariant measure on a certain manifold~${\Sigma_1}$ of matrices, selected
by optimal localization. The non pointwise nature of optimal localization
is the source for the breakdown of the covariance of \eqref{LQ}
under Lorentz
boosts\footnote{
Let ~$\sigma$ be a real antisymmetric matrix in the joint 
spectrum~$\Sigma$ of the~$Q^{\mu\nu}$'s; then there is a unique
irreducible 
representation~$i\lambda_P^2\sigma^{\mu\nu}=[q^\mu_\sigma,q^\nu_\sigma]$
of the DFR commutation relations by means of (non covariant)
quantum coordinates $q^\mu_{\sigma}$,
and all regular irreducible 
representations arise in this way. $\Sigma$ is
the orbit of the standard symplectic matrix~$\sigma_0=(\begin{smallmatrix}
0&-I\\I&0\end{smallmatrix})$  under the action of 
the full Lorentz group, while ${\Sigma_1}\subset\Sigma$ is the orbit
of~$\sigma_0$ under the rotation subgroup. 
See~\cite{dfr}.}. The 
Lagrangian~(\ref{LQ}) is an effective interaction obtained by considering
the quantization~$\wickdots{\phi(q)^n}$ of~$\wickdots{\phi(x)^n}$,
and was first proposed in~\cite{dfr}. 
We recall that the DFR 
quantization {\em \`a la Weyl} of $\phi(x)$ is
$\phi(q)=\int dk\;\check\phi(k)\otimes e^{ik_\mu q^\mu}$, where $\phi(x)$
is the ordinary neutral Klein--Gordon quantum field.

We also consider
\begin{equation}
\label{LE}
\mathscr L_I^E(x)=\Phi(x),
\end{equation}
where~$\Phi(x)$ is defined by
\[
\Phi(\bar q)=E\{\wickdots{\phi(q_1)\dotsm\phi(q_n)}\}.
\]
Here,~$E$ is the quantum diagonal map introduced in~\cite{bdfp}, which
generalizes the ordinary restriction 
$f\restriction_{\{x_j=x\}}=f(x,\dotsc,x)$ to the case of functions
$f(q_1,\dotsc,q_n)$ of~$n$ independent quantum events, 
so to respect the quantum limitations on 
localizability (once again, minimizing the differences $q_i-q_j$ in the
sense of optimal localization is responsible for the breakdown of full Lorentz 
covariance). 
Moreover,~$\bar q^\mu$ are the ``little'' 
quantum coordinates of the mean position\footnote{Note that the uncertainty
of a coordinate~$\bar q^\mu$ of the mean position 
of~$n$ independent 
quantum events is~$1/\sqrt n$ times that of a single event; by 
``little'', we mean that the joint spectrum of the operators
$-i[\bar q_\mu,\bar q_\nu]$ is~$(\lambda_P^2/n){\Sigma_1}$ instead of the full
manifold~$(\lambda_P^2/n)\Sigma$. 
The coordinates~$\bar q^\mu$ are not covariant under
Lorentz boosts. See~\cite{bdfp}.},  corresponding to ${\Sigma_1}$, and
$\Phi(\bar q)=\int dk\;\check \Phi(k)\otimes e^{ik_\mu\bar q^\mu}$ 
is the DFR quantization of~$\Phi(x)$ with respect to the coordinates~$\bar q$.

The interactions~$\mathscr L_I^Q$
and~$\mathscr L_I^E$  both 
are covariant under translations and space rotations. 
The Lagrangian~$\mathscr L_I^E$ was found ultraviolet regular in~\cite{bdfp}.

Notwithstanding that it fails to be covariant even under 
space rotations, the Lagrangian~$\mathscr L^\sigma_I$ 
has also been extensively 
investigated on its own in the literature, 
motivated either by~\cite{dfr} or by later results in string theory. 
Our remarks also apply to this class of models, even when
the matrix~$\sigma$ fails to be in~${\Sigma_1}$.  
In the latter case, however, the corresponding quantum coordinates
fail to fulfill the DFR stability conditions
(this is the case, e.g., of theories where the time coordinate commutes with 
all the space coordinates, $\sigma^{0j}=0$). 

The diagrammatic expansion of the Gell-Mann \& Low formula for such 
interactions has been extensively investigated by several authors 
(see e.g. \cite{sibold}). These and related questions were 
also carefully investigated in \cite{bahns}. The price to 
pay for the breakdown of locality apparently were the need for a more elaborate
topology of the diagrams than in the local case, and additional difficulties
in dealing with the time ordering; see also~\cite{bdfp}.

An important step was taken
in~\cite{schweda}, where the possibility of 
treating the time ordering by means of single diagrams (instead of taking a
different diagram for each arrangement of the ``time stamps'') was first
considered.

Here, we observe that the algebra of the Wick theorems for the
reduction of general time ordered products is essentially 
unaffected by the breakdown of locality. This allows a considerable
simplification in the computations, which are basically 
the same as in the local case. 
It easily follows from this simple remark that
the usual diagrams of the standard local perturbation theory 
can also be used in our case. We will obtain an alternative set of rules 
for ordinary diagrams, which contains the standard Feynman
rules for the usual local theory as a special case. 
This complements the discussion of \cite{bdfp0}. 

In the next section, we describe how to perform the Wick reduction of
general time ordered products. In section \ref{thema}, 
we use the general Wick theorems
to obtain the diagrammatic expansion of the Gell-Mann \& Low formula,
and we show that the vacuum-vacuum parts cancel out. In 
section \ref{variations}
we apply the framework to the motivating examples, and work out 
an example for the Lagrangian~(\ref{Lsigma}); in particular, 
we  recover the modified rules for the modified diagrams already considered in
the literature (we also shortly comment on the case where the Wick ordering
is suppressed in (\ref{L})). 
We finally draw some conclusions in section \ref{fugue}.

\section{General Wick Theorems.}

We consider the situation where there is a one-to-one correspondence 
between~$k$ points~$y_1,\dotsc,y_k$ of the space time and 
$k$ parameters~$\tau_1,\dotsc,\tau_k$, which we will call the times,
for short (not to be confused 
with the time variables appearing in the free fields
$\phi({y_j}^0,\by_j)$). Then, we can define the general time ordered product
\begin{equation}
\label{ordTprod}
\begin{split}
T^{\{\tau_j\}}&[\phi(y_1)\dotsm\phi(y_k)]=\\
=&\sum_\pi\phi(y_{\pi(1)})\dotsm\phi(y_{\pi(k)})
\theta(\tau_{\pi(1)}-\tau_{\pi(2)})\dotsm
\theta(\tau_{\pi(k-1)}-\tau_{\pi(k)}),
\end{split}
\end{equation}
which, for any choice of the points~$y_j$, coincides with the product
of the~$\phi(y_j)$'s taken in the order of decreasing times~$\tau_j$.
The sum runs over all permutations~$\pi$ of~$k$ elements. 

We make no assumptions on the~$\tau_j$'s: they may be constants, as well
as functions of the $y_j$'s, or even depend on extra parameters.

Then, we define a general chronological contraction {\em with respect to 
the given correspondence $y_j\leftrightarrow \tau_j$} as
\[
\begin{split}
&\wickdots{\phi(y_i)\dotsm\underwick{2}{<1\phi(y_i)\dotsw>1\phi(y_j)}
\dotsm\phi(y_n)}=\\
&=\wickdots{\phi(y_1)\dotsm\widehat{\phi(y_i)}\dotsw\widehat{\phi(y_j)}
\dotsm\phi(y_n)}\mathscr D(y_i-y_j;\tau_i-\tau_j),
\end{split}
\]
where a caret indicates omission, and 
\begin{equation}\label{eureka}
\mathscr D(x;\tau)=\frac{1}{i}\bigg(
\Delta_{\Pf}(x)\theta(\tau)+
\Delta_{\Pf}(-x)\theta(-\tau)\bigg)
\end{equation}
is the general propagator, which was called contractor in~\cite{schweda};
here we prefer to emphasize that Wick reduction works as usual, up to
putting ``general'' in front of everything. 

Indeed,
with these only modifications, the Wick reduction of 
$T^{\{\tau_j\}}[\phi(y_1)\dotsm\phi(y_k)]$ can be performed according to the
same rules as in the local case (``first Wick theorem''); in particular,
$T^{\{\tau_j\}}[\phi(y_1)\dotsm\phi(y_k)]$
equals the sum of the terms which can be obtained by applying all possible
choices of any number of general 
contractions (including no contraction at all) to
$\wickdots{\phi(y_1)\dotsm\phi(y_k)}$.

The argument is exactly the same as in the original proof of 
Wick~\cite{wick}, which only relies on the fact 
that~$[\phi(x),[\phi_{\Pf}(x'),\phi_{\Nf}(x'')]]=0$;  
the particular choice of a rule by which the times $\tau,\tau',\tau''$ are 
associated to $x,x',x''$
plays no role.  Here, we will content ourselves 
with checking the first Wick theorem in the simplest case:
consider
\[
T^{\{\tau,{\tau}'\}}[\phi(y)\phi(y')]
=\phi(y)\phi(y')\theta(\tau-{\tau}')+
\phi(y')\phi(y)\theta({\tau}'-\tau);
\]
replacing 
\begin{align*}
\phi(y)\phi(y')&=\wickdots{\phi(y)\phi(y')}+\frac{1}{i}\Delta_{\Pf}(y-y'),\\
\phi(y')\phi(y)&=\wickdots{\phi(y')\phi(y)}+\frac{1}{i}\Delta_{\Pf}(y'-y),
\end{align*}
in the above expression, 
and using~$\wickdots{\phi(y)\phi(y')}=\wickdots{\phi(y')\phi(y)}$, 
one obtains
\[
T^{\{\tau,{\tau}'\}}[\phi(y)\phi(y')]=
\wickdots{\phi(y)\phi(y')}+\mathscr D(y-y';\tau-\tau')=
\wickdots{\phi(y)\phi(y')}+
\wickdots{\underwick{2}{<1\phi(y)>1\phi(y')}},
\]
indeed.

The ordinary time ordering and Wick theorems are reobtained as a 
special case, setting the time~$\tau$ corresponding to~$y$ equal
to the time component~$y^0$ of~$y$ itself. Indeed, as already observed
in~\cite{schweda},
\[
\mathscr D(x;x^0)=\Delta_{SF}(x)
\]
is the usual Stueckelberg-Feynman propagator~\cite{stueckelberg,feynman}.

Since, by definition, $\wickdots{\phi(x)}=\phi(x)$, then (\ref{ordTprod}) is a 
special case of the general Wick ``mixed product'', where we closely follow the
original terminology of Wick. We find it convenient to introduce
the compact notations
\begin{equation}
\label{notations}
\uy=(y^1,\dotsc,y^{r}), \quad\phi^{(r)}(\uy)=\phi(y^1)\dotsm
\phi(y^r),
\end{equation}
so that the general Wick mixed product is defined by
\begin{equation}
\label{genWmix}
\begin{split}
T^{\{\tau_j\}}&[\wickdots{\phi^{(r_1)}(\uy_1)}\dotsm
\wickdots{\phi^{(r_k)}(\uy_k)}]=\\
=&\sum_\pi
\wickdots{\phi^{(r_{\pi(1)})}(\uy_{\pi(1)})}\dotsm
\wickdots{\phi^{(r_{\pi(k)})}(\uy_{\pi(k)})}\;
\theta(\tau_{\pi(1)}-\tau_{\pi(2)})\dotsm
\theta(\tau_{\pi(k-1)}-\tau_{\pi(k)});
\end{split}
\end{equation}
here all the points $y_j^1,\dotsc,y_j^{r_j}$ belonging to the $j^{\text{th}}$
monomial $\wickdots{\phi^{(r_j)}(\uy_j)}$ are made to correspond to the
same time parameter $\tau_j$. 

The Wick reduction of a general Wick mixed product can be obtained 
similarly to the preceding case, where now the contractions between fields 
belonging to different Wick monomials are forbidden (``second Wick theorem'').

See appendix \ref{wick-app} for a check of the second general Wick theorem
in the simplest case, 
and the review~\cite{p} for the general proofs of all the above statements. 

Of course, as in the ordinary case, matrix elements
of general time ordered products
may fail to make sense even as (tempered, say) distributions, and may
well need renormalization; the existence
of general time ordered products should be explicitly discussed
for each choice of the times~$\tau_j$. Even when they exist, finite
renormalization may well be necessary to perform the large scale limit.

\section{Theme.}\label{thema}

Let 
\[
S=I+\sum_{N=1}^\infty \frac{i^N}{N!}
\int dt_1\dots dt_N\;T[H_I(t_1)\dotsm H_I(t_N)]
\] 
be the Dyson series defined by the interaction Hamiltonian 
$H_I(t)$~\cite{dyson}, where $T$ indicates the standard Dyson time ordering
prescription. We consider the Gell-Mann \& Low formula
\begin{equation}
\label{additionalGL}
G_k(x_1,\dotsc,x_k)=\frac{\Langle T[\phi(x_1)\dotsm\phi(x_k)S]\Rangle_0}%
{\langle S\rangle_0},
\end{equation}
for the Green functions of the interacting fields~\cite{gml}, 
where $\langle\cdot\rangle_0$ is the expectation on the free (Fock) vacuum,
and the $\phi(x_j)$'s are free fields; we recall that here
the Dyson time ordering prescription is extended so that, after 
expanding $S$ in \ref{additionalGL}, at each order $N$ the product of the 
operators $\phi(x_1),\dotsc,\phi(x_k),H_I(t_1),\dotsc,H_I(t_N)$ is taken in 
the order of decreasing ${x_1^0},\dotsc,{x_k}^0, 
t_1,\dotsc,t_N$.

Following \cite{dfr}\footnote
{See in particular the end of \cite[Sect.\ 6]{dfr}.
The Dyson series and the Gell-Mann \& Low formula 
are purely quantum mechanical; 
the standard derivation of~(\ref{Gpert}) is not affected by the breakdown
of locality and causality. Non locality remains hidden in the definition
of the Lagrangean density.} we replace
the Hamiltonian $H_I(t)=-g\int d\bx\; \mathscr L_I(t,\bx)$, 
where $\mathscr L_I$
is of the form~(\ref{L}), in the Gell-Mann \& Low 
formula,
\begin{equation}
\label{Gpert}
\begin{split}
G_k(x_1,&\dotsc,x_k)=\frac{1}{\langle S\rangle_0}
\sum_{N=0}^\infty\frac{(-ig)^N}{N!}\int dx_{k+1}\dotsm dx_{k+N}\\
&\big\langle
T[\phi(x_1)\dotsm\phi(x_k)\mathscr L_I(x_{k+1})\dotsm \mathscr L_I(x_{k+N})]
\big\rangle_0
\end{split}
\end{equation}
and we obtain
\begin{equation}
\label{GGpert}
\begin{split}
G_k(x_1,&\dotsc,x_k)=\\
=&\frac{1}{\langle S\rangle_0}\sum_{N=0}^\infty
\frac{(-ig)^N}{N!}\int dx_{k+1}\dotsm dx_{k+N}
\int d\uy_{1}\dotsm d\uy_{N}\;W_{x_{k+1}}(\uy_1)\dotsm W_{x_{k+N}}(\uy_N)\\
&\big\langle
T^{\{{x_j}^0\}}[\phi(x_1)\dotsm\phi(x_k)
\wickdots{\phi^{(n)}(\uy_1)}\dotsm\wickdots{\phi^{(n)}(\uy_N)}]\big\rangle_0,
\end{split}
\end{equation}
where we use the notations (\ref{notations}).

We can now reduce the general
mixed time ordered product which appears in~(\ref{GGpert}),
where the times driving the time ordering are~$\tau_j={x_j}^0$ for 
$\protect{j=1,\dotsc,k+N}$.

If a general contraction involves two external fields
$\phi(x_i),\phi(x_j)$, then the general
propagator is~$\Delta_{SF}(x_i-x_j)$; if it involves an external field
$\phi(x_i)$ and a field~$\phi(y_j^u)$ in a Wick monomial, then the
general propagator is
\[
\mathscr D(x_i-y_j^u;{x_i}^0-{x_j}^0)
\]
if it involves  two fields~$\phi(y_i^u)$ and~$\phi(y_j^v)$ belonging
to distinct Wick monomials, then we pick a
\[
\mathscr D(y_i^u-y_j^v;{x_i}^0-{x_j}^0);
\]
the remaining contractions are forbidden.

Due to the form of~(\ref{L}), the kernel~$W_x$ can be assumed to be 
totally symmetric in~$y_1,\dotsc,y_n$, for any~$x$ fixed (otherwise,
we could symmetrize it without changing the interaction). One easily checks
that, with the rules \ref{FR1}--\ref{FR4} given below, 
\[
G_k(x_1,\dotsc,x_k)=\frac{1}{\langle S\rangle_0}\sum_\gamma 
m(\gamma) I_\gamma(x_1,\dotsc,x_k),
\]
where the sum runs over all the 
diagrams with~$k$ external points and any number
of unlabeled vertices\footnote{\label{diagfoot}
We label the external points by~$x_1,\dotsc,x_k$, but we
use diagrams with {\em unlabeled} vertices; this is the same as labeling 
the vertices, while declaring that, anyway, 
\[
\feyn{\vertexlabel^{x_1} f\vertexlabel^{y_1\;\;\;}}\feyn{fl}
\feyn{\vertexlabel^{\;\;\;y_2} f\vertexlabel^{x_2}}\;\;=\;\;
\feyn{\vertexlabel^{x_1} f\vertexlabel^{y_2\;\;\;}}\feyn{fl}
\feyn{\vertexlabel^{\;\;\;y_1} f\vertexlabel^{x_2}}\;.
\] 
Correspondingly, each diagram $\gamma$ has multiplicity
$v(\gamma)!m(\gamma)$. Note that if $\gamma$ arises at order $N$ in the 
perturbation series, then $N!=v(\gamma)!$.}. 
With~$v(\gamma)$ the number of vertices of~$\gamma$, 
\begin{equation}
\label{mgamma}
m(\gamma)=\frac{(n!)^{v(\gamma)}}{s(\gamma)};
\end{equation}
above, \(s(\gamma)\) is the product of \(\ell!\) over 
all unordered pairs of vertices, where \(\ell\) is the number of lines 
connecting each pair.

The contribution $I_\gamma$ corresponding to the diagram $\gamma$
(external points labeled by~$x_1,\dotsc,x_k$) is given by the integral
of the expression obtained by means of the following rules: 
\renewcommand{\theenumi}{R\arabic{enumi}}
\renewcommand{\labelenumi}{\theenumi}
\begin{enumerate}
\item
\label{FR1} 
{\em label each vertex by an index~$j$ in~$\{k+1,\dotsc,k+v(\gamma)\}$,
and otherwise arbitrary (different indices for different vertices);
for each vertex~$j$, take a factor 
\begin{equation}\label{vertex}
-igW_{x_j}(y_j^1,\dotsc,y_j^n)dx_jdy_j^1\dotsm dy_j^n;
\end{equation}}
\item
\label{FR2}  
{\em for each line connecting two external points, take as a factor
the corresponding  Stueckelberg-Feynman propagator;}
\item
\label{FR3} 
{\em for each line connecting the external point~$x_i$ and the
$j^{th}$ vertex, pick at random one of the unused~$y_j^1,\dotsc,y_j^n$,
say~$y_j^u$, then take 
\[
\mathscr D(x_i-y_j^u;{x_i}^0-{x_j}^0)
\]
and mark~$y_j^u$ as used;}
\item
\label{FR4}  
{\em for each line connecting the~$i^{\text{th}}$ and~$j^{\text{th}}$
vertices, pick at random two unused~$y_i^u$ and~$y_j^v$, then
take
\[
\mathscr D(y_i^u-y_j^v;{x_i}^0-{x_j}^0)
\]
and mark~$y_i^u, y_j^v$ as used.}
\end{enumerate}

The adiabatic switch and/or the restriction to finite volume 
are/is obtained as usual by turning the coupling constant
$g$ into a suitable function~$g(x_j)$ in rule~\ref{FR1}. 

If we take $W_x(y_1,\dotsc,y_n)=
\prod_j\delta(y_j-x)$, then (\ref{L}) is the ordinary local 
Lagrangian~$\wickdots{\phi(x)^n}$.
By integrating the variables~$y_j$ out of (\ref{vertex}), 
and correspondingly setting 
$y_j^u=x_j$ in rules~\ref{FR2},\ref{FR3}, we obtain the 
usual Feynman rules of the local perturbation theory as a special case.

It is straightforward to check that~$I_\gamma$ factorizes over the 
components of~$\gamma$, so that in particular one can factorize the
vacuum-vacuum parts. Since the combinatorics is the same as 
in the local case (the combinatorial factors do not depend on the choice
of the kernels), it is natural to expect that
the vacuum-vacuum parts cancel precisely with
$1/\langle S\rangle_0=\Big(\sideset{}{_\gamma^{(0)}}
\sum m(\gamma)I_\gamma\Big)^{-1}$ 
by the same mechanism as in the local case 
(the sum  $\sum_\gamma^{(0)}$ is restricted to the vacuum-vacuum diagrams, 
i.e.\ the diagrams with no external points).
Indeed, by inspection of (\ref{mgamma}), 
one finds that \(m(\gamma)\)
factorizes over the connected 
components of $\gamma$; since the most general diagram \(\gamma\) is the
disjoint union of a vacuum-vacuum diagram and a diagram with no 
vacuum-vacuum components, we obtain
\begin{equation}
\label{GMLnonloc}
G_k(x_1,\dotsc,x_k)=\sideset{}{^{(1)}}
\sum_\gamma m(\gamma)I_\gamma(x_1,\dotsc,x_k),
\end{equation}
where the sum $\sum_\gamma^{(1)}$ 
is restricted to the diagrams with no vacuum-vacuum components.
We thus reproduced in the present setting 
the original argument of the early days of local 
perturbation theory.

Equations representing the Fourier transform 
$\widehat I_\gamma(p_1,\dotsc,p_k)$
of $I_\gamma(x_1,\dotsc,x_k)$ are not so manageable; see however 
\cite{schweda}. In the local case, the Stueckelberg--Feynman 
propagators are functions of differences of points;
their Fourier transforms as functions of two variables then contain delta 
functions which make the convolutions trivial. Here this nice feature 
is lost for lines connecting an external point to a vertex; moreover, 
as a function of the difference $y_i^u-y_j^v$, the
general propagator $\mathscr D(y_i^u-y_j^v;{x_i}^0-{x_j}^0)$
connecting two vertices has Fourier transform
depending parametrically on other variables; the resulting expression is rather
cumbersome. One might observe that 
$\mathscr D(x-x';\tau-\tau')$ depends on the difference  $(x,\tau)-
(x',\tau')$ of 5-vectors; however, we refrain from pursuing this idea,
since it does not seem to provide any substantial simplification. 

\section{Variations.}\label{variations}
The above rules are suitable for general discussions. On practical purposes,
however, it may be useful to specialize the rules to the kernel which is
actually considered, also for the sake of comparison with the existing
literature. 

\subsection{First Variation.}
For the Lagrangian~(\ref{LE}), the kernel~$W_x$ is of the form
\begin{equation}
\label{exot}
W_x(y_1,\dotsc,y_n)=\delta\big(x^{0}-\tau(y_1,\dotsc,y_n)\big)
w(y_1,\dotsc,y_n),
\end{equation}
where both~$\tau$ and~$w$ are totally symmetric. See~\cite{bdfp}
for explicit expressions;~$\tau$ is called there the ``average time'' 
of the Lagrangian. This allows for integrating
out the~${x_j}^0$'s ($j> k$). 
Correspondingly, we replace (\ref{vertex}) by
\[
-igw(y_j^1,\dotsc,y_j^n)d\bx_jdy_j^1\dotsm
dy_j^n;
\]
in rule~\ref{FR1}, and we replace each time argument ${x_j}^0$  corresponding
to a vertex (i.e. $j>k$) by~$\tau(\uy_j)$ in the general propagators of
rules~\ref{FR3},\ref{FR4}. 

The coupling constant
$g$ has to be turned into a function 
$g(\tau(\uy_j),\bx_j)$, to investigate the adiabatic
switch and/or the finite volume theory.

\subsection{Second Variation.}
For~$n$ even, the kernel~$W_x$ 
for the Lagrangian\footnote{Here,  
$W_x(y_1,\dotsc,y_n)=\int_{{\Sigma_1}}d\sigma\;C(y_1-x,\dotsc,y_n-x)$,
where the explicit form of the kernel~$C_n$ is given in
eq.s (C.2) and (C.4) of~\cite[App. C]{dfr} (replace $Q$ by $\sigma$ in 
those equations).} 
(\ref{LE}) does not 
appear to have a compact expression for its totally symmetric part.
Hence each integral has to be expanded into~$m(\gamma)$ distinct integrals.
In this case, it is perhaps more practical to let the diagrams take care
of the symmetrization; moreover, one might wish to adhere to the philosophy
that each diagram should correspond to one integral. This also will explain the
reason why diagrams with more elaborate topologies may arise in some contexts.

\begin{figure*}
\begin{centering}
\epsfig{file=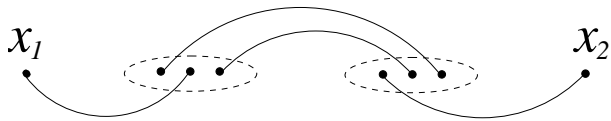, height=1.5cm}
\caption{Fat diagram (in position space) of 
a second order contribution to 
$G_2(x_1,x_2)$; the points surrounded by a dashed line belong to the
same fat vertex, and are ordered from left to right. Dashed lines only are a 
device to indicate sets of points, no topological information is attached to 
them. This connected diagram is of class
$\;\;\Feyn{\vertexlabel^{x_1}fflf\vertexlabel^{x_2}}\;\;$.}
\label{fatdiagram} 
\end{centering}
\end{figure*}
We consider ``fat diagrams'' (see 
figure \ref{fatdiagram}): a fat diagram~$\bgamma$ consists
of external points and ``fat vertices''; a fat vertex
is an {\em ordered} set of n points; each point originates exactly
one line; no line can connect two points in the same fat vertex; 
fat vertices are unlabeled; two diagrams
are the same if they can be made to coincide (by means of a
smooth deformation in 3-space)
so that the inner order of the points within each fat vertex is respected.
These diagrams are essentially those already considered in the literature
(see \cite{bdfp,schweda}).

We may introduce an equivalence relation on fat diagrams; two fat diagrams
$\bgamma,\bgamma'$ are equivalent if they shrink to the same ordinary
diagram~$\gamma$ 
when we shrink the fat vertices to points (equivalently: if they are the same
fat diagrams up to the inner order in the fat vertices). 
In this case, we say that 
$\bgamma,\bgamma'$ both are of class~$\gamma$. By definition,
a fat diagram~$\bgamma$ is connected if it is of class~$\gamma$ with~$\gamma$
connected (equivalently: we say that two points in a fat diagram 
are connected by a chain of 
lines if discontinuities only take place within fat vertices; then~$\bgamma$
is connected if any two of its points are connected by a chain of lines).

With these modifications,
\[
G_k(x_1,\dotsc,x_k)=\sideset{}{^{(0)}}
\sum_{\bgamma}\frac{1}{s(\gamma)}J_{\bgamma}(x_1,\dotsc,x_k),
\] 
where \(s(\gamma)=(n!)^{v(\gamma)}/m(\gamma)\), and 
we replace the rules~\ref{FR3},\ref{FR4} by the 
rules~\ref{FR3var},\ref{FR4var}
 below; note that
rule~\ref{FR1} remains unchanged up to considering ``vertex'' as a
synonym of ``fat vertex'', 
but now~$W_x$ may fail to be symmetric.
Rules~\ref{FR3var},\ref{FR4var} differ from~\ref{FR3},\ref{FR4}
in that we keep track
of the inner order of each fat vertex, by associating the~$u^{\text{th}}$
variable~$y_j^u$ in~$(y_j^1,\dotsc,y_j^n)$ to the~$u^{\text{th}}$
point in the~$j^{\text{th}}$ fat vertex.

\renewcommand{\theenumi}{R\arabic{enumi}'}
\renewcommand{\labelenumi}{\theenumi}
\begin{enumerate}
\addtocounter{enumi}{2}
\item
\label{FR3var}
{\em For each line connecting the external point~$x_i$ to the 
$u^{\text{th}}$ point in the 
$j^{th}$ vertex, take 
the factor
\[
\mathscr D(x_i-y_j^u;{x_i}^0-{x_j}^0);
\]}
\item
\label{FR4var}
{\em  for each line connecting the~$u^{\text{th}}$ point in the 
$i^{\text{th}}$ vertex to the~$v^{\text{th}}$ point in the~$j^{\text{th}}$
vertex, take the factor
\[
\mathscr D(y_i^u-y_j^v;{x_i}^0-{x_j}^0).
\]}
\end{enumerate}
Note that~$J_{\bgamma}$ also factorizes over the connected components
of~$\bgamma$.
With~$[\bgamma]$ the equivalence class of~$\bgamma$, we write 
$[\bgamma]=\gamma$ to say that~$\bgamma$ is of class~$\gamma$. Then
setting 
\[
I_\gamma(x_1,\dotsc,x_k)=\frac{s(\gamma)}{m(\gamma)}
\sum_{\bgamma\in\gamma}J_{\bgamma}(x_1,\dotsc,x_k),
\]
we recover the general expansion~(\ref{GMLnonloc}).
Note also that, if~$W_y$ is totally symmetric, then~$I_\gamma=J_{\bgamma}$
for each~$\bgamma$ of class~$\gamma$.

It is possible to extend 
the methods of Wick reduction of the Gell-Mann \& Low formula to the case
where the Wick ordering in (\ref{L}) is suppressed.
Note that, in that case, the kernel $W_x$ cannot be symmetrized,
so that we should resort to the fat diagrams of the present
variation. Moreover, lines connecting different points $y_j^u,y_j^v$
within the same 
fat vertex (oriented according to the inner order of fat vertices, namely from
$y_j^u$ to $y_j^v$ if $u<v$), should be allowed in the fat diagrams; 
these additional lines would correspond to the ordinary two--points 
functions\footnote{To see this in a simple example, we
replace $\phi(y)\phi(z)=
\wickdots{\phi(y)\phi(z)}+\tfrac{1}{i}\Delta_{\Pf}
(y-z)$ into 
$\phi(x)\phi(y)\phi(z)\theta(\tau-\tau')+
\phi(y)\phi(z)\phi(x)\theta(\tau'-\tau)$, obtaining 
$\phi(x)\tfrac{1}{i}\Delta_{\Pf}(y-z)+T^{\{\tau,\tau'\}}
[\phi(x)\wickdots{\phi(y)\phi(z)}]$;
by Wick reduction of the chronological product, the latter expression equals
$\phi(x)\tfrac{1}{i}\Delta_{\Pf}(y-z)+
\phi(y)\mathscr D(x-z;\tau-\tau')+\phi(z)\mathscr D(x-y;\tau-\tau')$.}
 $\tfrac{1}{i}\Delta_{\Pf}(y_j^u-y_j^v)$. 
This way, one could recover the rules discussed in \cite{schweda}.

\subsection{Third Variation.}
For~$n$ odd, the kernel~$W_x$ corresponding to the Lagrangian
(\ref{LQ}) is of the form~(\ref{exot}), where none
of~$w,\tau$ is symmetric. Hence one could  mix the flavours of the first
and second variation, to write down the appropriate rules. 
We leave this to the reader, and we work out an example instead,
using the variant~(\ref{Lsigma}) of the Lagrangian (no integration over
$\sigma$).

We consider
the second order contribution to the two points function~$G_2(x_1,x_2)$
with non local~$\phi^3$ self interaction,
corresponding to the fat diagram~$\bgamma$ 
(of class $\;\;\Feyn{\vertexlabel^{x_1}fflf\vertexlabel^{x_2}}\;\;$)
of figure  \ref{fatdiagram}.
Using the kernels\footnote{With~$Q$ replaced by~$\sigma$. Note that they only
are  meaningful when $\sigma$ fulfill~$\det\sigma\neq 0$, which is not the case
of time/space commutativity. Explicit kernels for the case $\det\sigma=0$
do not seem to be available in the literature, to the best of the author's
knowledge.} 
computed in~\cite[Appendix C]{dfr},
we get
\[
H_I(t)=-\frac{1}{\pi^4}\int dy_1dy_2dy_3
\wickdots{\phi(y_1)\phi(y_2)\phi(y_3)}
e^{2i(y_1-y_3)\sigma^{-1}(y_2-y_3)}\delta^{(1)}(t-\tau),
\]
where
\[
\tau(y_1,y_2,y_3)={y_1}^0-{y_2}^0+{y_3}^0,
\]
and~$x\sigma^{-1} y=x^\mu (\sigma^{-1})_{\mu\nu} y^\nu$.
The desired integral is then
\[
\begin{split}
J_{\bgamma}(x_1,x_2)=&
\frac{(ig)^2}{\pi^8}\int \bigg(\prod_{u=1}^3dy_1^udy_2^u\bigg)
e^{2i(y^1_1-y^3_1)\sigma^{-1}(y^2_1-y^3_1)+
2i(y^1_2-y^3_2)\sigma^{-1}(y^2_2-y^3_2)}\\
&\mathscr D(x_1-y_1^2;{x_1}^0-\tau(\uy_1))
\mathscr D(x_2-y_2^1;{x_2}^0-\tau(\uy_2))\\
&\mathscr D(y_1^1-y_2^3;\tau(\uy_1)-\tau(\uy_2))
\mathscr D(y_1^3-y_2^2;\tau(\uy_1)-\tau(\uy_2)).
\end{split}
\]

\section{Conclusions.}\label{fugue}
We obtained a unified treatment of the Wick reduction of time ordered products
and of the diagrammatic expansion of the 
Gell-Mann \& Low formula for the non local Lagrangians of interest to us;
the latter is based on the usual diagrams, and contains the ordinary local 
Feynman rules as a special case. We also recovered from it the modified rules 
for the modified diagrams which are currently investigated in the literature.

Indeed, the non local Lagrangians considered here are defined in terms of 
a length scale: the Planck length~$\lambda_P\sim 1.6\times 10^{-33}\text{cm}$.
In the large scale limit, where the quantum nature of the spacetime is
not directly visible any more, we expect to get back 
``diagramwise'' to the ordinary (non renormalized) local~$\phi^n$ model.

In principle, it should be possible
to implement in position space a natural variant of the
BPHZ renormalization procedure
\cite{bogopara,hepp,zimmermann}; 
the $R$-operation is a rather general
method for subtracting divergences, whose mechanism is quite insensitive to
the particular prescription for the subtractions. Moreover, it might be 
possible to adapt the reformulation of the BPHZ 
procedure in terms of Hopf algebras, due to Kreimer \cite{kreimer}.

Of course, a very  delicate point
is to detect the correct subtraction prescription. 

We
will discuss these issues in a forthcoming paper, under a different,
equivalent point of view.

\appendix
\section{The Second General Wick Theorem.}
\label{wick-app}

According to the second general Wick theorem, the general Wick mixed 
product (\ref{genWmix}) 
equals the sum of the terms 
obtained by applying all possible choices of any number (including none)
of allowed general contractions to
\[
\wickdots{\;\overbrace{\phi(y_1^1)\phi(y_1^2)\dotsm\phi(y_1^{r_1})}^{\tau_1}\;
\overbrace{\phi(y_2^1)\phi(y_2^2)\dotsm\phi(y_2^{r_2})}^{\tau_2}\;\dotsm\;
\overbrace{\phi(y_k^1)\phi(y_k^2)\dotsm\phi(y_k^{r_k})}^{\tau_k}\;}\;,
\]
where no contraction is allowed, which involves two fields associated with the 
same $\tau_j$. 

Since, once again, the proof is the same as in \cite{wick}, 
we check this statement for the purpose of illustration 
in the case of the mixed product
\begin{equation}
\label{3mixed}
\begin{split}
T^{\{\tau,\tau'\}}&[\phi(y)\wickdots{\phi(y_1')\phi(y_2')}]=\\
&=\phi(y)\wickdots{\phi(y_1')\phi(y_2')}\theta(\tau-\tau')+
\wickdots{\phi(y_1')\phi(y_2')}\phi(y)\theta(\tau'-\tau).
\end{split}
\end{equation}
We have ($\phi^{\PNf}$ are the creation and annihilation parts, as usual)
\begin{eqnarray*}
\lefteqn{\phi(y)\wickdots{\phi(y_1')\phi(y_2')}=}\\
&=&\wickdots{\phi^{\Pf}(y)\phi(y_1')\phi(y_2')}+\phi^{\Nf}(y)
\wickdots{\phi(y_1')\phi(y_2')}=\\
&=&\wickdots{\phi(y)\phi(y_1)\phi(y_2')}+\frac{1}{i}
\Delta_{\Pf}(y-y_1')\phi(y_2')
+\frac{1}{i}\Delta_{\Pf}(y-y_2')\phi(y_1).
\end{eqnarray*}
Analogously,
\begin{eqnarray*}
\lefteqn{\wickdots{\phi(y_1')\phi(y_2')}\phi(y)=}\\
&=&\wickdots{\phi(y_1)\phi(y_2')\phi(y)}+
\frac{1}{i}\Delta_{\Pf}(y_1'-y)\phi(y_2')
+\frac{1}{i}\Delta_{\Pf}(y_2'-y)\phi(y_1).
\end{eqnarray*}
Replacing the above in~(\ref{3mixed}), we obtain
\begin{eqnarray*}
\lefteqn{T^{\{\tau,\tau'\}}[\phi(y)\wickdots{\phi(y_1')\phi(y_2')}]=}\\
&=&\wickdots{\phi(y)\phi(y'_1)\phi(y_2')}
+\mathscr D(y-y_1')\phi(y_2')+\mathscr D(y-y_2')\phi(y'_1)=\\
&=&\wickdots{\phi(y)\phi(y'_1)\phi(y_2')}+
\wickdots{\underwick{2}{<1\phi(y)>1\phi(y'_1)\phi(y_2')}}
+\wickdots{\underwick{2}{<1\phi(y)\phi(y'_1)>1\phi(y_2')}};
\end{eqnarray*}
note that the (ill defined) general 
contraction $\wickdots{\underwick{1}{\phi(y)
<1\phi(y'_1)>1\phi(y_2')}}$ is missing.

\noindent{\bf acknowledgments.} I am grateful to Sergio Doplicher for his 
constant support, the  
many stimulating conversations, and his precious suggestions. 
It is also my pleasure to acknowledge the fruitful 
correspondence with Dorothea Bahns, and to thank Gerardo Morsella for
his comments on a preliminary version.

\end{document}